\documentstyle[12pt]{article}
\begin{document}

\begin{center}
{\bf QUANTUM DYNAMICS WITH INTERMEDIATE MEASUREMENTS 
     IN AGREEMENT WITH THE CLASSICAL DYNAMICS }

\vspace{1.2cm} B.\,KAULAKYS

{\em Institute of Theoretical Physics and Astronomy,}

{\em A.\,Go\v stauto 12, 2600 Vilnius, Lithuania} 
\vspace{1cm}

ABSTRACT \vspace{0.2cm}

\parbox{5in}
{\small The effect of repetitive measurement for quantum
dynamics of driven by an intensive external force of the
simple few-level systems as well as of the multilevel
systems that exhibit the quantum localisation of classical
chaos is investigated. Frequent measurement of the simple
system yields to the quantum Zeno effect while that of the
suppressed quantum system, which classical counterpart
exhibits chaos, results in the delocalisation of the
quantum suppression. From the analysis we may conclude
that continuously observable quasiclassical system evolves
essentially classically-like.}
\end{center}

\vspace{0.8cm}

\noindent {\bf 1. Introduction }\vspace{0.4cm}

Dynamics of a quantum system, when it is not being
observed, may be described by the Scr\"odinger equation,
by the quantum Liouville equation for the density matrix
or by the equation for evolution of the Wigner function.
In general, the quantum and classical realms are related
by the correspondence principle: physical characteristics
of the highly exited quantum systems with large quantum
numbers are close to those of its classical counterpart.
However, recent researches have suggested that the
use of the correspondence principle for strongly driven
nonlinear systems is not so straightforward: a quantum
interference effect suppresses the classical diffusion-like
chaotic motion and results in the interference
of macroscopic systems' states. The exponential instability
of classical motion characterised by the positive maximal
Lyapunov exponent destroys the deterministic image of the
classical physics and results in the unpredictability of 
the trajectory. Such stochasticity implies a continuous
spectrum of the motion. On the other hand, the frequency
spectrum of any quantum system, which motion is bounded
in the phase space, is always discrete. Accordingly,
the motion of such a system is regular. Therefore,
the question: can (and if yes, how) quantum mechanics
give chaos as a limiting behaviour, is open until now [1].

It is postulated in the von Neumann axiomatics of the
quantum mechanics that any measurement changes abruptly
the systems under consideration state and projects it to an
eigenstate of the measured observation. The measurement
process follows the irreversible dynamics and results
to the disappearance of coherence of the system's state:
to the decay of the off-diagonal matrix elements of the
density matrix or randomization of the phases of the wave
function's amplitudes.

Quantum system undergoes relatively slow evolution at an
early period after preparation or measurement [2].
Therefore, the repetitive frequent observation of the
quantum system can inhibit the decay of unstable or
dynamics of the driven by an external driven field system.
This phenomenon is called the quantum Zeno effect [3].
It should be noticed, that until now the quantum Zeno
effect has mostly been intensively analysed for the purely
quantum systems consisting only of the few (usually two
or three) quantum states. However, it is of interesting
to investigate the influence of the repetitive frequent
measurement on the evolution of the multilevel
quasiclassical systems, the classical counterparts
of which exhibit chaos. It has been established that
chaotic dynamics of such systems, e.g. dynamics of
strongly driven by periodic external field nonlinear
systems, is suppressed of the quantum interference effect
and results in the quantum localisation of the classical
dynamics in the energy space of the system
(see, e.g. [4-5]). Thus, the quantum localisation
phenomenon strongly limits the quantum motion of the
unobservable system.

As it was mentioned above, the repeated frequent 
measurement or continuous observation of quantum system 
can freeze its dynamics too. It is natural to expect 
that frequent measurement of the suppressed system will 
result in the additional freezing of the system's state. 
However, as we will see late, such supposition is wrong.

Here we show how the quantum Zeno effect in a 
two-level-system may be described by the wave function 
without the density matrix formalism. Then we use such 
a method for analysis of dynamics of the multilevel 
systems affected by repeated measurement. We reveal 
that the repetitive measurement of the multilevel 
systems with quantum suppression of classical chaos
results in the delocalisation of the states 
superposition and acceleration
of the chaotic dynamics.\vspace{0.8cm}

\noindent {\bf 2. Dynamics of the two-level systems }
\vspace{0.4cm}

The simplest time evolution of the two-state wave 
function 
$\Psi =a_1\left|1\right\rangle +a_2\left| 
2\right\rangle$ 
from time moment $t_k=k\tau$ to 
$t_{k+1}=(k+1)\tau$ may be represented as%
$$
\left( \matrix{a_1(k+1)\cr a_2(k+1)}\right) =
{\bf A}\left(
\matrix{a_1(k)\cr a_2(k)}\right) ,\eqno(2.1) 
$$
$$
{\bf A}=\left( 
\matrix{\cos\varphi& i\sin\varphi\cr 
i\sin\varphi & \cos\varphi}\right) ,~~~\varphi =
{\frac 12}\Omega \tau \eqno(2.2) 
$$
where $\Omega $ is the Rabi frequency. Evidently 
the evolution of the amplitudes from time $t=0$ 
till $t=n\tau$ may be expressed as%
$$
\left( \matrix{a_1(n)\cr a_2(n)}\right) =
{\bf A}^n\left(\matrix{a_1(0)\cr 
a_2(0)}\right) . \eqno(2.3) 
$$
One can calculate matrix ${\bf A}^n$ by the method 
of diagonalization of the matrix ${\bf A.}$ The result 
naturally is%
$$
{\bf A}^n=\left(
\matrix{\cos n\varphi & i\sin n\varphi\cr
i\sin n\varphi & \cos n\varphi}\right) .\eqno(2.4) 
$$
For the time interval $T=n\tau =\pi /\Omega$ a certain 
(with the probability $1$) transition between the 
states takes place.

Measurement of the system's state in the time moment 
$t=k\tau$ projects the system to the state 
$\left| 1\right\rangle$ with the probability 
$p_1(k)=\mid a_1(k)\mid ^2$ or to the state 
$\left| 2\right\rangle $ with the probability 
$p_2(k)=\mid a_2(k)\mid ^2$. After each of the 
measurement the phases of the amplitudes $a_1(k)$ 
and $a_2(k)$ are random which results in the absence 
of the interference terms in the expressions for the
probabilities. The evolution from the time $t=0$ until 
$t=n\tau$ with the $(n-1)$ intermediate measurement 
is described by the equation%
$$
\left( \matrix{p_1(n)\cr p_2(n)}\right) =
{\bf M}^n\left( \matrix{p_1(0)\cr 
p_2(0)}\right) . \eqno(2.5) 
$$
Matrix ${\bf M}^n$ calculated by the diagonalization 
method is%
$$
{\bf M}^n={\frac 12}\left( 
\matrix{1+\cos ^n2\varphi & 1-\cos ^n2\varphi\cr
1-\cos ^n2\varphi & 1+\cos ^n2\varphi}\right) .\eqno(2.6)
$$
From eqs. (2.5) and (2.6) we get the quantum Zeno 
effect [3]: if initially the system is in state 
$\left| 1\right\rangle $, the outcome of evolution
until the time $T=n\tau =\pi /\Omega$ with the 
intermediate measurements will be given by the 
probabilities 
$p_1(T)=(1+\cos ^n2\varphi )/2\rightarrow 1$ and 
$p_2(T)=(1-\cos ^n2\varphi )/2\rightarrow 0,~~(n
\rightarrow \infty )$%
. This result represents the inhibition of the quantum 
dynamics by measurements and confirms the proposition 
that act of measurement may be expressed as randomisation 
of the amplitudes' phases. \vspace{0.8cm}

\noindent {\bf 3. Dynamics of the multilevel systems }
\vspace{0.4cm}

In general the Schr\"odinger equation for strongly 
driven multilevel systems can not be solved analytically. 
However, the mapping form of quantum equations of motion 
greatly facilitates investigation of stochasticity and
quantum--classical correspondence for the chaotic 
dynamics. From the standpoint of an understanding of 
manifestation of the measurements for the dynamics of the 
multilevel systems the region of large quantum numbers 
is of greatest interest. Here we may use the quasiclassical 
approximation with the convenient variables angle $\theta $ 
and action $I$. The simplest system in which the dynamical
chaos and quantum localisation of states may be observed
is a system with one degree of freedom described by the
nonlinear Hamiltonian, $H_0(I)$, and driven by the periodic
$V(\theta ,t)=k\cos \theta \sum\limits_j\delta (t-j\tau )$
kicks [4-5]. Integrating the Schr\"odinger equation over
the period $\tau$ we obtain a map [5-7]%
$$
a_m(t_{j+1})=e^{-i\beta _m}\sum\limits_na_n(t_j)
J_{m-n}(k),~~\beta
_m=H_0(m)\tau ,~~t_j=j\tau \eqno(3.1)
$$
for the amplitudes $a_m(t_j)$ before the appropriate
$j$-kick in expansion of the state function
$\Psi (\theta ,t)$ through the eigenfuntions,
$\varphi_m=i^{-m}e^{im\theta }/\sqrt{2\pi }$,
of the action $I=-i\frac \partial{\partial \theta }$.
Here $J_m(k)$ is the Bessel function.

Quantum dynamics represented by map (3.1) with the
non-linear Hamiltonian $H_0(I)$ is similar to the
classical one only for some finite time $t^{*}$
after which it reveals an essential decrease of the
diffusion rate asymptotically resulting in the
exponential localisation of the system's state with
the localisation length $\lambda \sim k^2/2$ [1,4,5].

Each measurement of the system's state between
$(j-1)$ and $j$ kicks projects it to one of the state
$\varphi _m$ with the probability
$P_m(t_j)=\left| a_m(t_j)\right| ^2$. After such
a measurement the phase of the amplitude $a_m(t_j)$
is random. Therefore, the influence of the measurements
for further dynamics of the system may be expressed as
replacement of the amplitudes $a_m(t_j)$ by the
amplitudes $\exp \left[ i2\pi g_m(t_j)\right] a_m(t_j)$,
where $g_m(t_j)$ is a random number in case of
measurement of the $\varphi _m$-state's population
before the $j$ kick and equals zero in absence of such
a measurement. So we may analyse the influence on the
dynamics of measurements performed after every kick,
after every $N$ kicks or of the measurements just of
some states, e.g. only of the initial state, and observe
the reduction of the quantum localisation effect in a
degree depending on the extent and frequency of the
measurement [7]. In the case of measurement of all states
after every kick we have the uncorrelated transitions
between the states and diffusion-like motion with
the quantum diffusion coefficient in the $n$-space%
$$
B(n)={\frac 1{2\tau }}\sum\limits_m(m-n)^2J_{m-n}^2(k)=
{\frac{k^2}{4\tau }} \eqno(3.2)
$$
which coincides with the classical one. Therefore,
the quantum evolution of frequently observable chaotic
system is more classical-like than dynamics of
the isolated system.

To facilitate the comparison between quantum and
classical dynamics it is convenient to employ the Wigner
representation, $\rho _W\left( x,p,t\right) $%
, of the density matrix . The Wigner function of the
system evolves according to equation
$$
\frac{\partial \rho _W}{\partial t}=
\left\{ H,\rho _W\right\} _M\equiv
\left\{ H,\rho _W\right\} +
\sum\limits_{n\geq 1}\frac{\hbar ^{2n}\left(
-1\right) ^n}{2^{2n}\left( 2n+1\right) !}
\frac{\partial ^{2n+1}V}{\partial
x^{2n+1}}\frac{\partial ^{2n+1}\rho _W}
{\partial p^{2n+1}} , \eqno{(3.3)}
$$
where by $\left\{ ...\right\} _M$ and $\left\{ ...\right\}$
are denoted the Moyal and the Poisson brackets,
respectively, while the Hamiltonian of the system is of
the form $H=p^2/2m+V\left( x,t\right)$. The terms in
eq. (3.3) containing Planck's constant and higher
derivatives give the quantum corrections to the classical
dynamics generated by the Poisson brackets. In the region
of regular dynamics one can neglect the quantum corrections
for very long time if the characteristic actions of the
system are large. For classically chaotic motion the
exponential instabilities lead to the development of the
fine structure of the Wigner function and exponential
growth of its derivatives. As a result, the quantum
corrections become significant after relatively short
time even for macroscopic bodies [8]. The extremely
small additional diffusion-like terms in eq. (3.3), which
reproduce the effect of interaction with the environment
or frequent measurement, prohibits development of the
Wigner function's fine structure and removes barriers
posed by classical chaos for the correspondence
principle.

\vspace{0.8cm}

\noindent {\bf 4. Conclusion}\vspace{0.4cm}

The quantum-classical correspondence problem caused
of the chaotic dynamics is closely related with the
old problem of measurement in quantum mechanics.
Even the simplest detector follows irreversible
dynamics due to the coupling to the multitude of vacuum
modes which results in the randomisation of the quantum
amplitudes' phases, decay of the off-diagonal matrix
elements of the density matrix or to the smoothing of
the fine structure of the Wigner distribution
function--what we need to obtain the classical equations
of motion. The repetitive measurement of the multilevel
systems with quantum suppression of classical chaos results
in delocalisation of the states superposition and
acceleration of the chaotic dynamics which is opposite to
the quantum Zeno effect in driven systems. In the limit of
the frequent full measurement or unpredictable interaction
with the environment the quantum dynamics of such
quasiclassical systems approaches the classical motion.

\vspace{0.8cm}

{\bf References} \vskip\baselineskip

\begin{enumerate}
\item  G. Casati and B. Chirikov, in {\it Quantum Chaos:
Between Order and Disorder,} ed. G. Casati and B. V. Chirikov,
(Cambridge University, 1994) p.3. \vspace{-0.5\baselineskip}

\item  L. A. Khalfin, {\it Zh. Eksp. Teor. Fiz.} {\bf 33}
(1958) 1371 [{\it Sov. Phys. JETP} {\bf 6} (1958) 1503];
J. Swinger, {\it Ann. Phys.} {\bf 9} (1960) 169.
\vspace{-0.5\baselineskip}

\item  B. Misra and E. C. G. Sudarshan, {\it J. Math. Phys.}
{\bf 18} (1976) 756; R. J. Cook, {\it Phys. Scr.} {\bf T21}
(1988) 49; W. M. Itano {\it et al}, {\it Phys. Rev}.
A {\bf 41} (1990) 2295; V. Frerichs and A. Schenzle,
{\it Phys. Rev. }A {\bf 44} (1991) 1962.
\vspace{-0.5\baselineskip}

\item  G. Casati, B. V. Chirikov, D. L. Shepelyansky,
and I. Guarneri, {\it Phys. Rep.} {\bf 154} (1987) 77;
F. M. Izrailev, {\it ibid} {\bf 196} (1990) 299.
\vspace{-0.5\baselineskip}

\item  V. Gontis and B. Kaulakys, {\it Sov. Phys.--Collec.:
Lit. Fiz. Sb.} {\bf 28}(6) (1988) 1. \vspace{-0.5\baselineskip}

\item  B. Kaulakys, in {\it Quantum Communications and
Measurement, }ed. V. P. Belavkin, O. Hirota and R. L. Hudson,
(Plenum Press, 1995) p.193. \vspace{-0.5\baselineskip}

\item  V. Gontis and B. Kaulakys, in {\it Proc. Intern.
Conf. on Nonlinear Dynamics, Chaotic and Complex Systems,}
(7-12 Nov. 1995, Zakopane, Poland). \vspace{-0.5\baselineskip}

\item  W. H. Zurek and J. P. Paz, {\it Phys. Rev. Lett.}
{\bf 72} (1994) 2508.
\end{enumerate}

\end{document}